\documentstyle[twocolumn,pre,aps,epsf]{revtex}
\begin{document}
\draft
%\narrowtext

%% FOR TWO COLUMN  ACTIVATE THE LINE BELOW
\twocolumn[\hsize\textwidth\columnwidth\hsize\csname @twocolumnfalse\endcsname

%\preprint{\today}
%{\bf
\title{
Comment on "Magnetic Breakdown at High Fields: 
Semiclassical and Quantum Treatments"
}

\author
{Keita Kishigi and Yasumasa Hasegawa}

\address
{Faculty of Science, Himeji Institute of Technology,
 Ako, Hyogo 678-1297, Japan}
\date{Received  18 September, 2000}
%\date{\today}
\maketitle
%\begin{abstract}
%\end{abstract}
\pacs{
PACS numbers: 71.18.+y, 71.20.Rv, 72.15.Gd
}

%% FOR TWO COLUMN ACTIVATE THE LINE BELOW
]

%\hspace{0.5cm}

\narrowtext

In a recent Letter,
Han {\it et al.}\cite{han} 
investigated the effects of finite temperature and 
 spin on  the de Haas-van Alphen (dHvA) 
oscillation in the two-dimensional model, where 
 the closed  orbit and open orbit exist.
%In their Fig. 6, they showed the $g$-factor dependence of 
%the Fourier transform amplitudes (FTAs) of 
%dHvA oscillation.
They obtained the effective mass from the fitting of  
the $g_e$-dependence of these FTAs    
 with the damping factor, 
\begin{eqnarray}
R_s&=&\cos (p \pi g_e m^* / 2 m_e),
\label{damping}
\end{eqnarray}
expected from the semiclassical description based on the
 Lifshitz-Kosevich formula\cite{Shoenberg84}.
In this Comment we argue that since the damping factor
  Eq.(\ref{damping}) is obtained for the system with fixed
chemical potential $\mu$, it is not
appropriate  for the system with 
the fixed electron number $N$, although 
the estimation of $m^*$ is the same. 
We also call attention that 
the spin
does not  reduce all FTAs but
enhances  some amplitudes  in the
system with magnetic breakdown. 
This interesting effect of spin 
\cite{kishigispin}
is seen in Fig. 6 of Ref.\cite{han}
but it was not mentioned.

For simplicity we first consider the two-dimensional free-electron system. 
In the case of fixed $\mu$, the magnetization  oscillates as a
`saw-tooth' given by\cite{Shoenberg84},
\begin{equation}
\frac{M}{\beta N_0}=- \sum_{p=1}^{\infty}
   \frac{\sin\left( 2 \pi p\left(\frac{\mu}{\beta H} -
    \frac{1}{2}\right) \right) }{\pi p},
\end{equation}
where $\beta=e\hbar/m^*c$ is a double Bohr magneton with effective
mass and $N_0$ is the total number of electrons at $H=0$. 
When the spin-splitting is taken into account, the dHvA oscillation is
just the sum of 
the contributions from  spin-up and spin-down electrons with the phase
shift of $\pm g_e m^* / 4 m_e$, as shown in Fig. 1(a),
resulting in the damping factor Eq.(\ref{damping}).
In the fixed $N$ case, on the other hand, the magnetization for 
$g_e=0$ is
given by \cite{Shoenberg84},
\begin{equation}
 \frac{M}{\beta N}=\sum_{p=1}^{\infty}
   \frac{\sin\left( 2 \pi p\frac{N}{2 \rho \beta H} 
     \right) }{\pi p},
\end{equation}
where $\rho$ is the density of states for each spin at $H=0$.
%The chemical potential moves with the Landau levels when the magnetic
%field varies. 
When the spin-splitting is considered, the chemical
potential jumps between a spin-up Landau level and a spin-down Landau level
 when $N/2 \rho \beta H$ is integer or integer plus $1/2$. As a
result, $\mp 2(g_e m^* / 2 m_e-[g_e m^*/2 m_e])$,
where $[g_e m/2m_0]$ is the integer part of $g_e m/2m_e$,
 is added to  $M/\beta N$ 
when $n<N/2  \rho \beta H < n+1/2$ or  $n +1/2 < N/2  \rho \beta H < 
n+1$, respectively, as shown in Fig. 1(b).
This modification of the magnetization is written by the damping
factor 
%for the $p$th harmonics 
as\cite{kishigispin,nakanospin}, 
\begin{eqnarray}
 R_{p}^{N} = 
 \left\{
 \begin{array}{cc}
   |1 -2(g_e m/2m_0-[g_e m/2m_0])| & {\rm for\ odd\ }p \\
   1                         & {\rm for\ even\ }p 
  \end{array}\right. .
\label{damping2}
\end{eqnarray}
Only for the fundamental frequency ($p=1$), the periods of the damping
factor with respect to $g_e$ for fixed $\mu$ (Eq. (\ref{damping})) and
fixed $N$ (Eq.(\ref{damping2})) are the same. 
The $g_e$-dependence of the peak amplitude  in Fig. 6 of
Ref.\cite{han} was fitted with the cosine curves (Eq.(\ref{damping})),
but it (especially for $f_{\alpha}$) can be fitted better with
Eq.(\ref{damping2}).
 
%
%
%%%%%%%%%%%%%%%%%%%%%%%%%%%%%%%%%%%%%%%%%%%%%%%%%%%%%%%%%%%%%%%%%%%%%%%
%%  Fig.1                                                            %%
%%%%%%%%%%%%%%%%%%%%%%%%%%%%%%%%%%%%%%%%%%%%%%%%%%%%%%%%%%%%%%%%%%%%%%%
\begin{figure}[t]
  \begin{center}
  %% \mbox{\psfig{figure=fig1.eps,width=7cm}}
  %% \epsfile{file=fig1.eps,width=7cm}
  \leavevmode \epsfxsize=6.5cm  \epsfbox{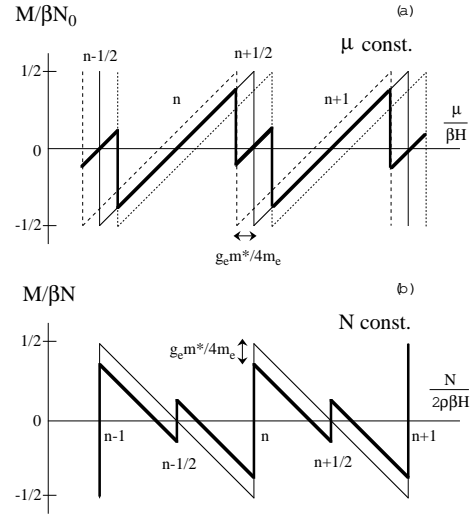}
  \end{center}
 \caption{
 Magnetization as a function of the inverse magnetic field.
(a) $\mu$ is fixed. (b) $N$ is fixed. Thin and thick lines are the
magnetization for $g_e=0$ and $g_e \not= 0$, respectively.   
 }
\end{figure}

The {\em enhancement} of the peak amplitude due to spin is seen
for the $f_{\beta+\alpha}$ and $f_{2\alpha}$ in Fig.6 of
Ref.\cite{han}. The enhancement of $f_{\beta+\alpha}$ has
been reported previously \cite{kishigispin}, 
where slightly simpler but  essentially the same model is
studied. 
We interpreted the enhancement of the amplitude as
the result of incomplete cancellation between the effects of the
chemical potential oscillation and magnetic breakdown\cite{kishigispin}.
%follows\cite{kishigispin}: 
%In the system where both magnetic breakdown and chemical potential
%oscillation takes place, the FTA for $f_{\beta+\alpha}$ is very small,
%while it is large in the system of fixed ${\mu}$ with magnetic
%breakdown or in the system with fixed $N$ without magnetic
%breakdown\cite{kishigispin,nakanospin}.  
%The chemical potential oscillation is reduced by the spin, but the
%magnetic breakdown will not  affected strongly. Then the cancellation
%of the effects of chemical potential oscillation and the magnetic
%breakdown becomes incomplete, and the FTA for  $f_{\beta+\alpha}$ is
%enhanced. 
We also observe the enhancement of the $\beta+2 \alpha$
oscillation\cite{kishigispin}.
The $g_e$-dependence of the $2 \alpha$ oscillation in Ref.\cite{han}
 is different from 
our result\cite{kishigispin}. This
difference may be due to the choice of  parameters in the calculations.

In summary, 
the damping factor (Eq.(\ref{damping})) which is commonly applied to
analyze the effects of spin,  should be
used carefully. In the system with magnetic breakdown, some Fourier
amplitudes are {\em enhanced} by  spin. 

%\hspace{.05cm}
%\author

%\noindent
%{Keita Kishigi and Yasumasa Hasegawa}
%
%%\address
%{Faculty of Science, Himeji Institute of Technology,
% Ako, Hyogo 678-1297, Japan}
%
%\noindent
%Received \today
%
%\noindent
%PACS numbers: 71.18.+y, 71.20.Rv, 72.15.Gd
%
%\vspace{-0.5cm}
%\begin{references}

%\end{references}


\begin{thebibliography}{99}

\vspace{-1.5cm}
\bibitem{han}
S. Y. Han, J. S. Brooks, and Ju H. Kim, Phys. Rev. Lett. {\bf 85}, 
1500 (2000).

%\bibitem{LK}
%I. M. Lifshitz and A. M. Kosevich, 
%Zh. Eksp. Theor. Fiz. {\bf 29}, 730 (1955);
%%Sov. Phys. JETP {\bf 2} (1956) 636;

\bibitem{Shoenberg84}
D.~Shoenberg, {\it Magnetic Oscillation in Metals}
(Cambridge University Press: Cambridge, 1984).

\bibitem{kishigispin}
K. Kishigi, Y. Hasegawa and M. Miyazaki, J. Phys. Soc. Jpn. {\bf 68}
 1817 (1999),
{\bf 69}, 821 (2000).


%\bibitem{kishigipaper}
%K. Kishigi, Y. Hasegawa and M. Miyazaki, J. Phys. Soc. Jpn. {\bf 69}, 
%821 (2000).

\bibitem{nakanospin}
M. Nakano: Phys. Rev. B {\bf 62}  45 (2000).

%\bibitem{kishigi2}
%K. Kishigi, M. Nakano, K. Machida, and Y. Hori: J. Phys. Soc. Jpn. {\bf
%64}
%(1995) 3043.

%\bibitem{kishigi3}
%K. Kishigi: J. Phys. Soc. Jpn. {\bf 66}
%(1997) 910.

\end{thebibliography}
\end{document}